\documentclass[12pt,preprint]{aastex}

\shorttitle{Collapsed Cores in Globular Clusters}
\shortauthors{Djorgovski et al.}

\newcommand{\lapprox}{{\footnotesize $\buildrel < \over \sim \,$}}
\newcommand{\gapprox}{{\footnotesize $\buildrel > \over \sim \,$}}

\begin{document}


\title{Shaping bipolar Planetary Nebulae : How mass loss leads to
waistline development.}


\author{C. Dijkstra \& A. K. Speck}
\affil{Department of Physics and Astronomy, University of Missouri,
Columbia, MO 65211, USA}
\email{dijkstrac@missouri.edu}






\begin{abstract}
Asymptotic Giant Branch (AGB) stars generally have spherically
symmetric envelopes, whereas most post-AGB stars and Planetary Nebulae
(PNe) show axisymmetric circumstellar envelopes. While various
mechanisms for axisymmetric circumstellar structures may explain the
shapes of PNe, they do not address how the shape of the circumstellar
shell evolves. Here we address the temporal changes in the axisymmetry
of AGB star envelopes, and in particular the development of the torus
required in the Generalized Interacting Stellar Winds (GISW)
model. Assuming (1) an AGB star rotates with sufficient angular speed
at the start of the AGB phase; and (2) that the rotational angular
momentum of the AGB star is conserved, we demonstrate that some very
important observational features of AGB star axisymmetry evolution can
be reproduced. We find that, compared to the star's increasing
luminosity and decreasing effective temperature, the decreasing mass
of the star primarily affects the axisymmetry of the envelope. When a
representative mass loss history is adopted, where most of the mass is
lost near the end of the AGB phase, the envelope's axisymmetry
increases over time, with the strongest increase occurring near the
end of the AGB phase. This may naturally explain why most AGB stars
have spherically symmetric envelopes, while axisymmetry seems
common-place in the post-AGB/PNe phase. The degree of axisymmetry at
the end of the AGB phase is found to increase with increasing main
sequence mass, and the onset of axisymmetry occurs only after the
onset of the superwind (SW) phase, in good agreement with the
observations.
\end{abstract}





\section{Introduction}

\label{sect:introduction}

Asymptotic Giant Branch (AGB) stars are the highly evolved descendants
of 1 \lapprox $M$ \lapprox 8\,M$_{\odot}$ main-sequence stars
\citep{1983ARA&A..21..271I}, which lose mass at a high rate
\citep[$10^{-7}$ to $10^{-4}{\ }\rm{M_{\odot}/yr}$; see
e.g.][]{1996A&ARv...7...97H,2004AGBstarsHabingOlofsson:C1}. The
outflowing matter creates a dusty molecular circumstellar envelope
which may completely obscure the central star. As the dust is driven
outwards by radiation pressure of the central star, it drags the
molecules in the envelope along with it. It is believed that mass-loss
rate increases significantly only towards the end of the AGB phase
\citep[e.g.][]{1993ApJ...413..641V,2002ApJ...581.1204V}. This increase
in mass-loss rate is necessary to explain the densities seen in
typical Planetary Nebulae \citep[PNe;][]{1981pprg.work..431R}. The
sudden and rapid increase in mass loss has been dubbed the {\it
superwind} \citep[SW;][]{1981pprg.work..431R}. Since the invocation of
the superwind, many observations of AGB stars have supported this
hypothesis
\citep[e.g.][]{1985ApJ...292..640K,1992ApJ...397..552W}. During the SW
phase the mass-loss rate, $\dot{M}$, exceeds
$\dot{M}_{\rm{SW}}\approx\rm{10^{-5}\,M_{\odot}/yr}$
\citep[e.g.][]{1981pprg.work..431R,1992ApJ...397..552W,2004AGBstarsHabingOlofsson:C3}. After
the SW phase AGB stars enter the post-AGB phase, or pre-planetary
nebula (PPN) phase, via which they may eventually evolve into
PNe. During the post-AGB/PN phase heavy mass loss stops, and the
envelope moves away from the star. Meanwhile, the central star heats
up and ionizes the circumstellar gas. The remains of the envelope
eventually disperse into the Interstellar Medium (ISM), while the
central star cools off as a white dwarf (WD).

AGB star evolution is intimately linked to its dust production. The
circumstellar dust drives mass loss and shapes the circumstellar
environments. While AGB stars generally have spherically symmetric
mass loss \citep[e.g.][]{2004AGBstarsHabingOlofsson:C7}, most post-AGB
stars and PNe show axisymmetric circumstellar envelopes
\citep{1999ApJS..122..221M,2004AGBstarsHabingOlofsson:C10,
2004ASPC..313..141S,2005AJ....130.2717S}. The axisymmetry is seen as a
bipolar or hourglass-shaped reflection nebula, where the two lobes
(bubbles, jets) of the nebula are separated by an equatorial waist, or
torus. The onset of the axisymmetry is believed to occur very near the
end of the AGB phase
\citep[e.g.][]{2004AGBstarsHabingOlofsson:C7,2004AGBstarsHabingOlofsson:C10}.
While some PNe are indeed spherical \citep{2005AJ....130.2717S}, these
appear to be old nebulae, and it is increasingly evident that the
majority of PNe are aspherical, and that most are bipolar
\citep{2005AJ....130.2717S}.

A suggested mechanism for shaping PNe is the Generalized Interacting
Stellar Winds (GISW) model
\citep[e.g.][]{1982ApJ...258..280K,1985MNRAS.212..837K}. In this model
a circumstellar torus is present in the equatorial plane of the AGB
star. The exact origin of the torus is not specified, but is agreed to
be somehow created by the slowly expanding AGB wind. Later, during the
post-AGB and PNe phase, a faster wind interacts with this torus. The
fast wind is blocked by the torus in the equatorial plane of the AGB
star, but may easily flow into the polar regions, creating a bipolar
(P)PN.

The development of axisymmetry, and in particular the origin of the
equatorial torus needed to create it in the GISW model, has been
attributed to various mechanisms. Here, the presence of a companion
star, fast rotating AGB stars, magnetic fields, and the compression of
gas into the equatorial region by two jets have been considered in
various ways. We will briefly discuss these mechanisms below.

A companion star may have several effects. First, it may cause the AGB
star to spin-up
\citep[e.g.][]{1994MNRAS.270..734H,2000MNRAS.317..861S}. This spin-up
may lead to small deviations from spherical symmetry of the AGB star
itself, which are subsequently amplified in the dust condensation
region due to the non-linear behavior and the strong temperature and
density dependence of the dust formation process
\citep{1996A&A...313..605D}. This then leads to preferential mass loss
along the equator. Second, to spin-up an AGB star, the companion must
be close (see Sect.\,\ref{sect:validityofassumptions} for a
quantification of the necessary proximity). If the companion is close
however, and the AGB star's mass loss rate is high enough, the
companion may accrete mass and create jets in the process. This also
leads to axisymmetry \citep[][and references
therein]{2005AJ....129..947S}. Finally, the gravitational influence of
a companion star may help shape the AGB star's envelope into an
axisymmetric geometry \citep{1998ApJ...497..303M,1999ApJ...523..357M}.

Several AGB stars and PNe show magnetic fields in their circumstellar
environments, and these fields are often suggested to be the main
agent responsible for creating axisymmetry. However,
\cite{2006PASP..118..260S} argues that a single star can not supply
the energy and angular momentum to create the large coherent magnetic
fields required for shaping the circumstellar wind, although magnetic
fields may have a secondary role.

Equatorially enhanced densities are commonly attributed to the
equatorially-enhanced mass loss of an AGB star. However, they may also
originate from the compression of gas into the equatorial region by
bipolar jets. \cite{2000ApJ...538..241S} argue that the interaction of
a slow AGB wind with a collimated fast wind (CFW) blown by a
main-sequence or white dwarf companion leads to equatorial density
enhancements. Here, the CFW originates from the accretion of the AGB
wind into a disk around the companion. The CFW forms two jets (lobes)
along the symmetry axis which compress the slow AGB wind near the
equatorial plane, leading to the formation of a dense slowly expanding
ring. Later, after the CFW and slow AGB wind cease, the primary star
leaves the AGB and blows a second, more spherical, fast wind. This
wind is then collimated by the dense equatorial material and leads to
a bipolar PN as described by the GISW model.

While these various mechanisms for axisymmetric circumstellar
structures may explain the shapes of PNe, they do not address the time
evolution of the shape of the circumstellar shell. In particular, they
generally fail to address the fact that most AGB shells are spherical
(and remain that way for most of the AGB phase), whereas most post-AGB
objects (and PNe) are bipolar (or multi-polar).

In this paper we address the temporal changes in the axisymmetry of
the circumstellar envelope, and in particular the development of the
torus that is required in the GISW model. For this purpose we consider
the scenario of an AGB star that rotates at a sufficient angular speed
at the start of the AGB phase. Next, as the star evolves, we assume
that the rotational angular momentum of the AGB star is somehow
conserved. Although simplistic, we will demonstrate that if the above
conditions are met, some very important observational features of AGB
star axisymmtery evolution may be reproduced. In
Sect.\,\ref{sect:modeldescription} we discuss our model and the
validity of its underlying assumptions. In
Sect.\,\ref{sect:resultsofthemodel} we discuss the results of the
model. Sect.\,\ref{sect:conclusions} lists our conclusions.

\section{Model description}

\label{sect:modeldescription}

We note that in order for what follows to succeed, the AGB star needs
to have a sufficiently large angular velocity at the start of the AGB
phase. This probably necessitates some sort of a spin-up, which will
be discussed in Sect.\,\ref{sect:axisymmetryandstellarevolution}.

The mass-loss rate is affected by the escape velocity. A rotating
sphere will tend to become oblate and will have a reduced
gravitational acceleration, $g$, at the equator compared to the
pole. With a lower $g$, the escape velocity is reduced and thus mass
loss increases. Therefore we need to consider the effect this will
have on the structure/shape of the circumstellar shell and how this
effect changes with time.

\subsection{Effective gravity ratio between the equator and poles}

We start by considering $\alpha$, which measures the ratio in
effective gravity between the equator and poles of a rotating AGB
star. For a test-mass $m$ at the surface of the star, it is given by
\begin{equation}
\alpha=\frac{GMm/R^2-m\omega^2R}{GMm/R^2}=1-\frac{\omega^2R^3}{GM}
\label{eq:Rgrav}
\end{equation}
where $G$ is the gravitational constant, and $M$ and $R$ are the mass
and radius of the star. $\omega$ represents the angular velocity of
the star at the equator. We next assume that the angular momentum
of the AGB star is conserved, i.e.
\begin{equation}
I\omega=I_{\rm{i}}\omega_{\rm{i}}
\label{eq:conservrotenerg}
\end{equation}
where $I$ represents the moment of inertia of the star. Here, we adopt
the notation that $x_{\rm{i}}$ indicates the initial value of a given
quantity $x$, i.e. the value of $x$ right after the spin-up. Assuming
solid-body rotation through the stellar envelope, and that at all
times the star's deviation from spherical symmetry is small, and
ignoring the core's moment of inertia and increasing mass during AGB
evolution, we have $I{\propto}M_{\rm{env}}R^{2}$, where $M_{\rm{env}}$
is the envelope's mass. From this it then follows that
\begin{equation}
\omega^{2}=\omega_{\rm{i}}^{2}\left(\frac{M_{\rm{env,i}}}{M_{\rm{env}}}\right)^{2}\left(\frac{R_{\rm{i}}}{R}\right)^{4}\quad.
\label{eq:omegasquared}
\end{equation}
Using Eq.\,\ref{eq:omegasquared} and
\begin{equation}
R=R_{\rm{i}}\sqrt{\frac{L}{L_{\rm{i}}}}{\left(\frac{T_{\rm{eff,i}}}{T_{\rm{eff}}}\right)}^{2}\quad,
\label{eq:Rstar}
\end{equation}
where $L$ and $T_{\rm{eff}}$ are the luminosity and effective
temperature of the star, we then find for $\alpha$
\begin{eqnarray}
\alpha&=&1-\frac{\omega_{\rm{i}}^2R_{\rm{i}}^3}{GM}\left(\frac{M_{\rm{env,i}}}{M_{\rm{env}}}\right)^2\sqrt{\frac{L_{\rm{i}}}{L}}\left(\frac{T_{\rm{eff}}}{T_{\rm{eff,i}}}\right)^{2}\nonumber\\
&=&1-\frac{\omega_{\rm{i}}^2R_{\rm{i}}^3}{GM}\left(\frac{M_{\rm{i}}-M_{\rm{c}}}{M-M_{\rm{c}}}\right)^2\sqrt{\frac{L_{\rm{i}}}{L}}\left(\frac{T_{\rm{eff}}}{T_{\rm{eff,i}}}\right)^{2}\quad,
\label{eq:Rgrav2}
\end{eqnarray}
where $M_{\rm{c}}$ is the core mass of the star.

Using Eq.\,\ref{eq:Rgrav2}, we may follow $\alpha$ as a function of
stellar evolution. Since the AGB star's deviations from spherical
symmetry are amplified in the dust condensation region
\citep{1996A&A...313..605D}, the behavior of the degree of axisymmetry
in the envelope may in principle be followed as a function of stellar
evolution as well.

\subsection{Axisymmetry of the envelope}

The ratio $\rho_{\rm{e}}/\rho_{\rm{p}}$, where $\rho_{\rm{e}}$ and
$\rho_{\rm{p}}$ represent the mass density distribution in the
equatorial and polar direction respectively, is a measure for the
degree of axisymmetry in the envelope. If the exact relation between
$\rho_{\rm{e}}/\rho_{\rm{p}}$ and $\alpha$ is known, and if we assume
that $\rho_{\rm{e}}/\rho_{\rm{p}}$ in first order only depends on
$\alpha$ (see below), we may monitor $\rho_{\rm{e}}/\rho_{\rm{p}}$
with stellar evolution.

While the relationship between $\rho_{\rm{e}}/\rho_{\rm{p}}$ and
$\alpha$ is non-trivial, we can still make an estimate for such a
relation. \cite{1996A&A...313..605D} modeled the structure of a
stationary dust driven wind for a $1\,\rm{M_{\odot}}$ AGB star
($T_{\rm{eff}}=2600$\,K, $L=1.0\times10^4\,L_{\odot}$ and
$1.2\times10^4\,L_{\odot}$) spun-up by a companion star to various
angular velocities, up to $2\times10^{-8}s^{-1}$ ($10^{-8}s^{-1}$) for
the $1.0\times10^4\,L_{\odot}$ ($1.2\times10^4\,L_{\odot}$)
star. Here, the spin-up leads to small deviations from spherical
symmetry of the AGB star itself, which are subsequently amplified in
the dust condensation region due to the non-linear behavior and the
strong temperature and density dependence of the dust formation
process. The result is an axisymmetric envelope. The models of
\cite{1996A&A...313..605D} provide rotationally-modulated solutions
for the structure of stationary dust-driven winds, both as a function
of rotation rate and polar angle. The output of their models includes
$\dot{M}_{\rm{e}}/\dot{M}_{\rm{p}}$ and $v_{\rm{e}}/v_{\rm{p}}$, the
mass-loss rate ratio and terminal outflow-velocity ratio between the
equatorial and polar direction.

\begin{figure}[!t]
\hspace{-4.5cm}
\includegraphics[angle=90,scale=.7]{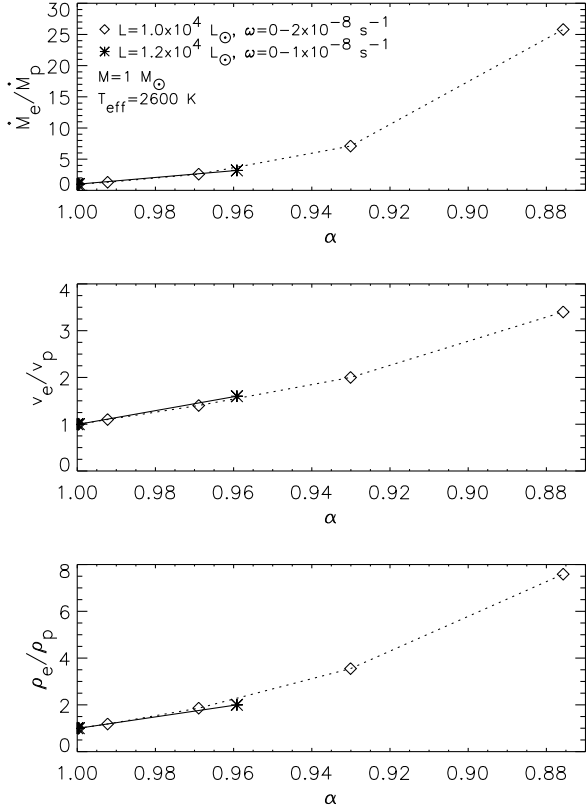}
\caption{$\dot{M}_{\rm{e}}/\dot{M}_{\rm{p}}$ (top),
$v_{\rm{e}}/v_{\rm{p}}$ (middle), and $\rho_{\rm{e}}/\rho_{\rm{p}}$
(bottom) as a function of $\alpha$. The curves were calculated from
detailed model results by \cite{1996A&A...313..605D}. The calculations
were done for a $1\,\rm{M_{\odot}}$, $T_{\rm{eff}}=2600$\,K, AGB star
for two different luminosities (for a $1.0\times10^4\,L_{\odot}$ star,
rotating up to $2\times10^{-8}s^{-1}$, and a
$1.2\times10^4\,L_{\odot}$ star, rotating up to $10^{-8}s^{-1}$).  For
details see the main text.}
\label{fig:figure1}
\end{figure}

Using Eq.\,\ref{eq:Rgrav} and
$\rho_{\rm{e}}/\rho_{\rm{p}}=(\dot{M}_{\rm{e}}/\dot{M}_{\rm{p}})/
(v_{\rm{e}}/v_{\rm{p}})$ we calculated $\alpha$ and
$\rho_{\rm{e}}/\rho_{\rm{p}}$ for these models, in order to estimate a
relation between them. The result is shown in
Fig.\,\ref{fig:figure1}. As expected, it can be seen that with
decreasing $\alpha$, i.e. stronger deviations from spherical symmetry
of the AGB star, $\rho_{\rm{e}}/\rho_{\rm{p}}$ increases, i.e. the
envelope becomes more axisymmetric. In the remainder of this paper we
will assume that this relation is valid for all AGB stars. The
validity of this assumption will be discussed below.

\subsection{Validity of assumptions}

\label{sect:validityofassumptions}

We now briefly discuss the validity of the assumption of conservation
of angular momentum of the star, and the assumption of the use of the
relation between $\alpha$ and $\rho_{\rm{e}}/\rho_{\rm{p}}$ found in
Fig.\,\ref{fig:figure1} for all AGB stars.

The most important assumption in our model is that after the spin-up,
the rotational angular momentum of the AGB star is conserved. The
validity of this assumption is uncertain, since there are various
possible sinks and sources of angular momentum for the star. Angular
momentum will be lost from the star through its stellar wind
\citep[e.g.][]{2000MNRAS.317..861S}. This loss can be substantial, and
may in principle remove all of the stars angular momentum. Ignoring
the magnetic influence beyond the stellar surface and the moment of
inertia of the core relative to that of the envelop, and assuming
solid body rotation through the envelope, \cite{2000MNRAS.317..861S}
showed that the amount of angular momentum in the envelope is
proportional to $J_{\rm{env}}\propto{M_{\rm{env}}}^{\delta}$, where
$\delta$ is a constant. For the upper AGB $\delta=3$. Using the
initial-final mass relation \citep{2000A&A...363..647W}, which links
the main sequence mass of a low or intermediate mass star with its
final mass after the AGB phase (i.e. its WD mass), and assuming that
the AGB phase terminates when the stellar envelope's mass has been
reduced to about $0.01\,\rm{M_{\odot}}$
\citep[e.g.][]{1995A&A...297..727B}, it can be shown that the envelope
mass of a $1\,\rm{M_{\odot}}$ ($7\,\rm{M_{\odot}}$) changes by a
factor $M_{\rm{env,i}}/M_{\rm{env}}=45\,(598)$, i.e. the angular
momentum of the star decreases by a factor
$J_{\rm{env,i}}/J_{\rm{env}}{\approx}\,9\times10^{4}(2\times10^{8})$\,!
The loss of angular momentum through the stellar wind is therefore
important. Other means by which the envelope may lose angular momentum
include magnetic breaking and expansion of the convective envelope on
the AGB \citep{1998A&A...334..210H}.

Sources of angular momentum for the envelope may be provided by
companion stars/planets and the core of the star. First, at the
expense of its orbital angular momentum, a low mass companion may feed
angular momentum to the envelope during a common envelope phase
\citep[e.g.][]{1994MNRAS.270..734H,2000MNRAS.317..861S}. The angular
momentum of a companion of mass $M_{\rm{p}}$ is given by
\citep{2000MNRAS.317..861S}
\begin{equation}
J_{\rm{p}}=8\times{10}^{49}\left(\frac{M_{\rm{p}}}{M_{\rm{J}}}\right)\sqrt{\frac{M_{\rm{i}}}{0.9\rm{M_{\odot}}}{\frac{a}{1\,\rm{AU}}}}\,\rm{g}\,\rm{cm}^2\,\rm{s}^{-1},
\label{eq:Jp}
\end{equation}
where $M_{\rm{J}}$ is Jupiter's mass and $a$ is the initial orbital
separation. Following \cite{2000MNRAS.317..861S}, we may express the
envelope's angular momentum as
\begin{eqnarray}
J_{\rm{env}}&=&I\omega\nonumber\\
&=&{\gamma}M_{\rm{env}}R^{2}\omega\nonumber\\
&\approx&2\times10^{52}\left(\frac{\gamma}{2/9}\right)\left(\frac{M_{\rm{env}}}{1\,\rm{M_{\odot}}}\right)\times\nonumber\\
&&\left(\frac{R}{1\,\rm{AU}}\right)^{2}\left(\frac{\omega}{10^{-8}\rm{s}^{-1}}\right)\,\rm{g}\,\rm{cm}^2\,\rm{s}^{-1}\quad,
\label{eq:Jenv}
\end{eqnarray}
where $\gamma$ is a constant ($\gamma=2/9$ is appropriate for the
upper AGB). Comparing Eq.\,\ref{eq:Jp} and Eq.\,\ref{eq:Jenv} shows
that e.g. a $M_{\rm{p}}\approx100\,M_{\rm{J}}=0.1\,\rm{M_{\odot}}$
companion star with an inital orbital separation of 1\,AU may replace
all of the rotational angular momentum lost by the envelope of a star
with $M_{\rm{i}}=1\,\rm{M_{\odot}}$,
$M_{\rm{env}}{\leq}0.45\,\rm{M_{\odot}}$, $R=1\,\rm{AU}$, and
$\omega=10^{-8}\rm{s}^{-1}$. Likewise, a
$M_{\rm{p}}\approx500\,M_{\rm{J}}=0.5\,\rm{M_{\odot}}$ companion star
with an inital orbital separation of 1\,AU may replace all of the
rotational angular momentum lost by the envelope of a star with
$M_{\rm{i}}=7\,\rm{M_{\odot}}$,
$M_{\rm{env}}{\leq}5.98\,\rm{M_{\odot}}$, $R=1\,\rm{AU}$, and
$\omega=10^{-8}\rm{s}^{-1}$. The above illustrates that (a substantial
fraction of) the angular momentum lost by the wind can in principle be
replaced by a low mass companion star.

Second, \cite{1999ApJ...517..767G} argue that although the envelope of
low- and intermediate-mass stars may be devoid of angular momentum
(i.e. non-rotating) at the beginning of the thermally pulsing AGB
phase, stars with main sequence masses \gapprox $1.3\,\rm{M_{\odot}}$
can spin up their envelopes to rotational speeds of
${\sim}1\,\rm{km/s}$ just prior to the PN ejection. Here, the spin-up
is caused by angular momentum from the core that {\it leaks} to the
envelope during thermal pulses \citep{1999ApJ...517..767G}.

The validity of our assumption of angular momentum conservation for
the star thus remains to be seen. However, as we will demonstrate in
the following section, if this condition is somehow met, some very
important observational features of AGB stars may be reproduced. It is
therefore interesting to consider the scenario where the star's
angular momentum is conserved, despite the uncertainty of this
assumption.

Strictly speaking, the relationship found between
$\rho_{\rm{e}}/\rho_{\rm{p}}$ and $\alpha$ (see
Fig.\,\ref{fig:figure1}, lower panel) is of course only valid for the
$1\,\rm{M_{\odot}}$ AGB star modeled above. Still, it may not be
unreasonable to assume that $\rho_{\rm{e}}/\rho_{\rm{p}}$ will in
first order primarily depend on $\alpha$, since it is the deviation
from axisymmetry of the AGB star that leads to an axisymmetric
envelope in the first place. If this assumption is correct, we may
apply the obtained relationship to other stars as well, and generally
study $\rho_{\rm{e}}/\rho_{\rm{p}}$ as a function of stellar
parameters and evolution (see below). If the assumption is invalid,
the behavior of $\alpha$ still provides information on the general
behavior of $\rho_{\rm{e}}/\rho_{\rm{p}}$ for a given star, since
deviations from spherical symmetry of the AGB star will still be
amplified in the dust condensation region
\citep{1996A&A...313..605D}. Only the exact relation between $\alpha$
and $\rho_{\rm{e}}/\rho_{\rm{p}}$ may not be specified in this case.

\section{Results of the model}

\label{sect:resultsofthemodel}

\subsection{Axisymmetry and stellar parameters}

As AGB stars evolve, their luminosity increases, while their mass and
effective temperature decrease. Following Eq.\,\ref{eq:Rgrav2}, the
effect of $L$ and $T_{\rm{eff}}$ is thus to increase $\alpha$
(i.e. lower the degree of axisymmetry) over time, while the effect of
$M$ will be to decrease $\alpha$ (i.e. increase the degree of
axisymmetry) over time. Here, the stellar mass has the largest effect
on $\alpha$ and hence the axisymmetry in the envelope. While the
luminosity of the star typically increases by a factor of 10 during
AGB star evolution, $\alpha$ (see Eq.\,\ref{eq:Rgrav2}) depends on
$\sqrt{L_{\rm{i}}/L}$, which only decreases by a factor of
3. Likewise, while $\alpha$ depends on
$(T_{\rm{eff}}/T_{\rm{eff,i}})^{2}$, the effective temperature
typically decreases from 3500\,K to 2500\,K, so
$(T_{\rm{eff}}/T_{\rm{eff,i}})^{-2}$ decreases at most by a factor
2. In constrast, $\alpha$ varies with the stellar mass as
$(1/M)((M_{\rm{i}}-M_{\rm{c}})/(M-M_{\rm{c}}))$. Assuming its tip of
the AGB envelope mass is $\sim0.01\,\rm{M_{\odot}}$
\citep{1995A&A...297..727B}, $\alpha$ may therefore vary by a factor
of about $3.6\times10^3$ ($2.5\times10^6$) over the lifetime of a
$1\,\rm{M_{\odot}}$ ($7\,\rm{M_{\odot}}$) star. The axisymmetry in the
envelope will thus indeed be primarily affected by the stellar mass.

\subsection{Axisymmetry and stellar evolution}

\label{sect:axisymmetryandstellarevolution}

\begin{figure}[!t]
\hspace{-0.825cm}
\includegraphics[angle=90,scale=.35]{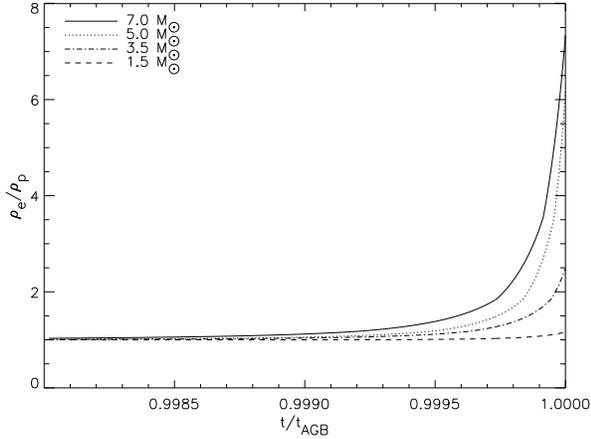}
\caption{$\rho_{\rm{e}}/\rho_{\rm{p}}$ as a function of time for stars
with initial masses of 1.5\,$\rm{M_{\odot}}$ ({\it dashed}),
3.5\,$\rm{M_{\odot}}$ ({\it dashed dotted}), 5.0\,$\rm{M_{\odot}}$ ({\it
dotted}), and 7.0\,$\rm{M_{\odot}}$ ({\it solid}). For details see the
main text.}
\label{fig:figure2}
\end{figure}

We now investigate the time evolution of $\rho_{\rm{e}}/\rho_{\rm{p}}$
under the influence of mass loss during the AGB phase. This is shown
in Fig.\,\ref{fig:figure2} for four stars with different main sequence
masses. It is assumed that each star is spun-up to an initial
rotational velocity of $0.076\,\rm{km/s}$. This choice of initial
rotational velocity ensures that, when combined with the other stellar
parameters adopted for each star, $\alpha$ is always in the range
covered by the models of \cite{1996A&A...313..605D}, therefore
allowing a translation to $\rho_{\rm{e}}/\rho_{\rm{p}}$ to be
made. After the spin-up each star has an initial luminosity of
$3000\,\rm{L_{\odot}}$, an initial effective temperature of 3500\,K,
an initial radius of $\sim0.7$\,AU, and an initial mass loss rate of
$\dot{M}_{\rm{i}}=10^{-7}\,\rm{M_{\odot}/yr}$. After spin-up, the
stars lose mass according to
\begin{equation}
\dot{M}(t)=\dot{M}_{\rm{i}}\times\exp{\left(\left(\frac{t}{t_{\rm{AGB}}}\right)^n{\ln{\left(\frac{\dot{M}_{\rm{f}}}{\dot{M}_{\rm{i}}}\right)}}\right)}\quad,
\label{eq:Mdot}
\end{equation}
where $\dot{M}_{\rm{f}}$ is the final mass loss rate, and the time $t$
is expressed in terms of the total AGB lifetime, $t_{\rm{AGB}}$. The
parameter $n$ may be found by specifying the onset of the SW phase,
$t_{\rm{SW}}$, which is also expressed in terms of
$t_{\rm{AGB}}$. Here we use
\begin{equation}
n=\frac{\ln(\ln(\frac{\dot{M}_{\rm{SW}}}{\dot{M}_{\rm{i}}})/\ln(\frac{\dot{M}_{\rm{f}}}{\dot{M}_{\rm{i}}}))}{\ln(\frac{t_{\rm{SW}}}{t_{\rm{AGB}}})}\quad,
\label{eq:n}
\end{equation}
where $t_{\rm{SW}}$ is estimated from detailed model results by
\cite{1993ApJ...413..641V}, and $t_{\rm{AGB}}$ and $\dot{M}_{\rm{f}}$
are adjusted such that the initial-final mass relation
\citep{2000A&A...363..647W} is satisfied for each star. Then the mass
of the star is found by
\begin{equation}
M(t)=M_{i}-\int^{t}_{0}\dot{M}(t')dt'\quad.
\label{eq:mass}
\end{equation}
The exact mass-loss rate as a function of time during the AGB phase is
unknown and difficult to determine. Still, a key feature of it seems
to be that most of the mass is lost near the end of the AGB phase
\citep[e.g.][]{1981pprg.work..431R,1993ApJ...413..641V}. With our
choice of $\dot{M}(t)$ this condition is satisfied. We may now insert
Eq.\,\ref{eq:mass} into Eq.\,\ref{eq:Rgrav2} to follow $\alpha$ as a
function of time. Next, we use the results of Fig.\,\ref{fig:figure1}
to translate $\alpha$ into $\rho_{\rm{e}}/\rho_{\rm{p}}$, yielding the
results shown in Fig.\,\ref{fig:figure2}. Note that we kept $L$ and
$T_{\rm{eff}}$ constant at $L=1{\times}10^{4}\,\rm{L_{\odot}}$ and
$T_{\rm{eff}}=2500$\,K in the above analysis (where $L$ and
$T_{\rm{eff}}$ are not to be confused with $L_{\rm{i}}$ and
$T_{\rm{eff,i}}$). As discussed earlier, the effect of $L$ and
$T_{\rm{eff}}$ are minor compared to the effects of the stellar
mass. These adopted values for $L$ and $T_{\rm{eff}}$ ensure that the
derived values for $\rho_{\rm{e}}/\rho_{\rm{p}}$ are in fact lower
limits. Table\,\ref{tab:table1} lists the parameters adopted for each
model in Fig.\,\ref{fig:figure2}.

\begin{table*}[!t]
\begin{center}
\setlength\tabcolsep{2.0pt}
\caption{Model parameters for the stars shown in
Fig.\,\ref{fig:figure2}. ${\Delta}t_{\rm{AX,2}}$ and
${\Delta}t_{\rm{AX,5}}$ represent the time before the end of the AGB
phase in which $\rho_{\rm{e}}/\rho_{\rm{p}}$ exceeds 2 and 5
respectively, i.e. when a reasonable axisymmetry may be observed in
our model calculations. The symbol $-$ is used when no value for
${\Delta}t_{\rm{AX,2}}$ or ${\Delta}t_{\rm{AX,5}}$ is available. For
an explanation of the other parameters see the main text.}
\vspace{0.2cm}
\begin{tabular}{|c|c|c|c|c|c|c|c|c|c|c|c|c|c|}
\tableline
$M_{\rm{i}}$  & $L_{\rm{i}}$  & $L$           & $T_{\rm{eff,i}}$ & $T_{\rm{eff}}$ & $R_{\rm{i}}$ & $v_{\rm{i}}$ & $\dot{M}_{\rm{i}}$ & $\dot{M}_{\rm{SW}}$ & $\dot{M}_{\rm{f}}$ & $t_{\rm{AGB}}$ & $t_{\rm{SW}}$\tablenotemark{\dagger}    & ${\Delta}t_{\rm{AX,2}}$ & ${\Delta}t_{\rm{AX,5}}$\\
($\rm{M_{\odot}}$) & ($\rm{L_{\odot}}$) & ($\rm{L_{\odot}}$) & (K)              & (K)            & (AU)         & (km/s)       & ($\rm{M_{\odot}}$/yr)   & ($\rm{M_{\odot}}$/yr)    & ($\rm{M_{\odot}}$/yr)   & (yr)           & ($t_{\rm{AGB}}$) & (yr) & (yr)\\
\tableline
1.5  & 3000 & 10000 & 3500 & 2500 & 0.7 & 0.076 & 1.0$\times10^{-7}$ & 1.0$\times10^{-5}$ & 3.0$\times10^{-5}$ & 2.3$\times10^{6}$ & 0.99 & --  & -- \\
3.5  & 3000 & 10000 & 3500 & 2500 & 0.7 & 0.076 & 1.0$\times10^{-7}$ & 1.0$\times10^{-5}$ & 3.0$\times10^{-5}$ & 2.1$\times10^{6}$ & 0.96 & 82  & -- \\
5.0  & 3000 & 10000 & 3500 & 2500 & 0.7 & 0.076 & 1.0$\times10^{-7}$ & 1.0$\times10^{-5}$ & 5.5$\times10^{-5}$ & 1.4$\times10^{6}$ & 0.92 & 202 & 23 \\
7.0  & 3000 & 10000 & 3500 & 2500 & 0.7 & 0.076 & 1.0$\times10^{-7}$ & 1.0$\times10^{-5}$ & 1.0$\times10^{-4}$ & 6.9$\times10^{5}$ & 0.80 & 162 & 30 \\
\tableline
\end{tabular}
\vspace{-0.6cm}
\tablenotetext{\dagger}{Values estimated from detailed models results by \cite{1993ApJ...413..641V}.}
\label{tab:table1}
\end{center}
\end{table*}

Fig.\,\ref{fig:figure2} shows that the axisymmetry increases over
time. Here, the strongest increase is clearly only near the very end
of the AGB phase, which is due to the fact that the star loses most of
its mass only at this stage
\citep{1993ApJ...413..641V,2002ApJ...581.1204V}. This result may
naturally explain why most AGB stars have spherically symmetric
envelopes, while in the post-AGB/PNe phase axisymmetry seems
common-place
\citep{1999ApJS..122..221M,2004AGBstarsHabingOlofsson:C10,2004ASPC..313..141S,2005AJ....130.2717S}.

Fig.\,\ref{fig:figure2} also shows that the final degree of
axisymmetry, i.e. the degree of axisymmetry at the end of the AGB,
increases with increasing main sequence mass.  This result also seems
to agree with observations. Indeed, an optical imaging survey of
post-AGB stars by \cite{2000ApJ...528..861U} with the Hubble Space
Telescope (HST) seems to suggest that higher mass progenitor AGB stars
are more likely to show bipolar structure with a completely or
partially obscured central star than low mass stars. This result seems
to be reinforced by detailed radiative transfer calculations of
\cite{2002ApJ...571..936M} for the post-AGB stars HD\,161796 and
IRAS\,17150-3224.

It is interesting to note that the onset of axisymmetry does not
coincide with the onset of the SW phase. This is illustrated in
Fig.\,\ref{fig:figure2} and Table\,\ref{tab:table1}. The onset of
axisymmetry only becomes reasonably appearant
(i.e. $\rho_{\rm{e}}/\rho_{\rm{p}}${\gapprox}2) in the last few tens
or hundreds of years, long after the start of the SW phase. This is in
agreement with e.g. recent findings by \cite{2006inprepUeta}. Based on
Spitzer Space Telescope \citep[SST;][]{2004ApJS..154....1W}
observations of NGC\,650 (Program ID\,77), \cite{2006inprepUeta} shows
that mass loss is nearly isotropic when the SW is turned on towards
the end of the AGB phase, and the SW mass loss precipitously decreases
along the polar direction during the SW phase.

The degree of axisymmetry that develops in the envelope will strongly
depend on the initial rotational velocity of the AGB star. Generally,
the rotation of single stars is expected to slow down significantly
while they are on the main-sequence
\citep[e.g.][]{1999ApJ...517..767G,2006PASP..118..260S}, and once
these stars reach the AGB phase they are believed to rotate too slowly
for axisymmetries to develop \citep{1996A&A...313..605D}. Here, slow
rotation is expected both with and without the presence of magnetic
fields \citep{2006PASP..118..260S}. It is therefore often argued that
a companion star is prerequisite to spin-up an AGB star
\citep[e.g.][]{1994MNRAS.270..734H,2000MNRAS.317..861S}. The mechanism
adopted in this paper would benefit from the presence of such a binary
companion, since it may spin-up the AGB star, and be a source of
rotational angular momentum for the AGB star's envelope that is
otherwise lost through the AGB wind (see
Sect.\,\ref{sect:validityofassumptions}). Moreover, the requirement of
a companion star would fit in nicely with the idea that binary
interaction indeed plays an important role in shaping the axisymmetric
structure of PNe (see Sect.\,\ref{sect:introduction}), and recent
observational results that strongly suggest that most or even all PNe
are in close binary systems \citep{2005AAS...207.9302D}.

Alternatively, in case of stars with main sequence masses \gapprox
$1.3\,\rm{M_{\odot}}$, the spin-up may be provided by angular momentum
from the core of the AGB star during thermal pulses (see
Sect.\,\ref{sect:validityofassumptions}), in which case a companion
star is not required. The thermal pulses will not provide the star
with a spin-up at the start of the AGB phase, since these pulses only
occur late in the AGB phase. However, they may efficiently spin-up the
AGB star near the end of the AGB phase \citep{1999ApJ...517..767G},
during which the mechanism adopted in our paper may still apply. In
this case, it seems more appropraite to use
$M_{\rm{env,i}}\approx0.1\,\rm{M_{\odot}}$, since at the onset of the
thermal pulses the mass of the envelope may already have been
substantially reduced down to $\sim0.1\,\rm{M_{\odot}}$
\citep{1999ApJ...517..767G}. Here, the lower initial envelope mass
will result in a lower degree of axisymmetry in the envelope, since
$\alpha$ (see Eq.\,\ref{eq:Rgrav2}) will stay closer to unity.

Although the trends described for $\rho_{\rm{e}}/\rho_{\rm{p}}$ are in
nice agreement with observations, the absolute values predicted for
$\rho_{\rm{e}}/\rho_{\rm{p}}$ are somewhat low. For example, we
predict a final value of $\rho_{\rm{e}}/\rho_{\rm{p}}\approx1.2$ for a
$1.5\,\rm{M_{\odot}}$ main sequence star. In contrast, for the
post-AGB star HD\,161796, which is believed to be of a low main
sequence mass \citep[e.g.][and references
therein]{2002ApJ...571..936M}, \cite{2002ApJ...571..936M} find
$\rho_{\rm{e}}/\rho_{\rm{p}}\approx9$. Also, we predict a final value
of $\rho_{\rm{e}}/\rho_{\rm{p}}\approx7.3$ for a $7.0\,\rm{M_{\odot}}$
main sequence star, which is on the high mass end of AGB stars. Still,
\cite{2002ApJ...571..936M} find for e.g. the high mass post-AGB star
IRAS\,17150-3224 $\rho_{\rm{e}}/\rho_{\rm{p}}\approx160$, which is
clearly much larger than predicted by our models. However, the trend
of an increasing value of $\rho_{\rm{e}}/\rho_{\rm{p}}$ with
increasing main sequence mass is the same.

There may be several reasons for our under-prediction of
$\rho_{\rm{e}}/\rho_{\rm{p}}$. First, the assumed initial rotational
velocity of $0.076\,\rm{km/s}$ is rather low. As discussed earlier,
this choice ensured that, when combined with the other stellar
parameters adopted for our stars, $\alpha$ is always in the range
covered by the models of \cite{1996A&A...313..605D}, therefore
allowing a translation to $\rho_{\rm{e}}/\rho_{\rm{p}}$ to be
made. This is a computational restriction however, and not a physical
one. In fact, rotational velocities higher than we adopted may be
expected \citep[see e.g.][and references
therein]{1999ApJ...517..767G}. Adopting a higher initial rotational
velocity will increase $\rho_{\rm{e}}/\rho_{\rm{p}}$. Also, we have
only considered the effects of mass thusfar. Taking properly into
account the effects of the luminosity and effective temperature of the
star as well may also help increase $\rho_{\rm{e}}/\rho_{\rm{p}}$. For
example, we adopted a luminosity of
$L=1{\times}10^{4}\,\rm{L_{\odot}}$ for all stars. Still, for stars
with low main-sequence masses this luminosity may be too
high. Lowering it will help increase $\rho_{\rm{e}}/\rho_{\rm{p}}$ in
this case. Finally, the values of $\rho_{\rm{e}}/\rho_{\rm{p}}$
derived by \cite{2002ApJ...571..936M} for HD\,161796 and
IRAS\,17150-3224 will depend on the details of their radiative
transfer calculations. Possibly, a different assumption of e.g. dust
composition or grain size distribution may lower the values of
$\rho_{\rm{e}}/\rho_{\rm{p}}$ derived by \cite{2002ApJ...571..936M},
in better agreement with our results.

\section{Conclusions}

\label{sect:conclusions}

In this paper we addressed the temporal changes in the axisymmetry of
AGB star envelopes, and in particular the development of the torus
that is required in the GISW model. For this purpose we considered the
scenario of an AGB star that rotates at a sufficient angular speed at
the start of the AGB phase. Next, as the star evolves, we assumed that
the rotational angular momentum of the AGB star is somehow conserved.
If the above condition is met, some very important observational
features of AGB star axisymmtery evolution may be reproduced. In
particular compared to the star's increasing luminosity and decreasing
effective temperature, the decreasing mass of the star primarily
affects the axisymmetry of the envelope. Adopting a representative
mass loss history, where most of the mass is lost near the end of the
AGB phase, results in an increase of the envelope's axisymmetry over
time. The strongest increase is only near the very end of the AGB
phase, explaining why most AGB stars have spherically symmetric
envelopes, while those of post-AGB stars and PNe are
axisymmetric. Also, the degree of axisymmetry at the end of the AGB
phase is found to increase with increasing main sequence mass, and the
onset of axisymmetry only occurs after the onset of the SW phase. Our
findings are in good agreement with observations.

\acknowledgments

This work was supported by the NASA Astrophysical Data Program (NAG
5-12675) and the University of Missouri Research Board. We are very
grateful to Noam Soker for directing our background reading regarding
the effects of binaries on the morphology of circumstellar envelopes,
and to Adam Frank and Toshiya Ueta for initial feedback. AKS would
like to thank her WS06 Stellar Astrophysics class for their indulgence
in debating the idea behind this paper during our discussion of
stellar evolution.

\end{document}